\documentclass[aip, amsmath,amssymb,
 reprint,%
nofootinbib, preprintnumbers
]{revtex4-1}

\usepackage{graphicx}
\usepackage{dcolumn}
\usepackage{bm}
\usepackage{lipsum}

\usepackage[utf8]{inputenc}
\usepackage[T1]{fontenc}
\usepackage{mathptmx}
\usepackage{etoolbox}

\makeatletter
\def\@email#1#2{%
 \endgroup
 \patchcmd{\titleblock@produce}
  {\frontmatter@RRAPformat}
  {\frontmatter@RRAPformat{\produce@RRAP{*#1\href{mailto:#2}{#2}}}\frontmatter@RRAPformat}
  {}{}
}%
\makeatother
\begin{document}


\title[]{Ultrafast and highly collimated radially polarized photons from a colloidal quantum dot in a hybrid nanoantenna at room-temperature}

\author{Alexander Nazarov$^\dagger$}
\affiliation{Racah Institute of Physics, The Hebrew University of Jerusalem, Jerusalem 9190401,Israel}
\email{ronenr@phys.huji.ac.il}

\author{Yuval Bloom$^\dagger$}%

\affiliation{Racah Institute of Physics, The Hebrew University of Jerusalem, Jerusalem 9190401,Israel}%

\author{Boaz Lubotzky}%

\affiliation{Racah Institute of Physics, The Hebrew University of Jerusalem, Jerusalem 9190401,Israel}%

\author{Hamza Abudayyeh}%

\affiliation{Department of Physics, University of Texas at Austin, Austin, 78712, Texas, USA}%

\author{Annika Mildner}%

\affiliation{Institute for Applied Physics and Center LISA$^+$, University of Tuebingen, Auf der Morgenstelle 10, 72076, Tuebingen, Germany}%

\author{Lorenzo Baldessarini}%

\affiliation{Department of Physics, Trento University, Via Sommarive, 14, Povo, 38123 TN, Italy}%

\author{Yuval Shemla}%

\affiliation{Racah Institute of Physics, The Hebrew University of Jerusalem, Jerusalem 9190401,Israel}%

\affiliation{Department of Physics, Columbia University, New York, 10027, New York, USA}%

\author{Eric G. Bowes}

\affiliation{Materials Physics \& Applications Division: Center for Integrated Nanotechnologies, Los Alamos National Laboratory,
Los Alamos, New Mexico 87545, USA}

\author{Monika Fleischer}

\affiliation{Institute for Applied Physics and Center LISA$^+$, University of Tuebingen, Auf der Morgenstelle 10, 72076, Tuebingen, Germany}

\author{Jennifer A. Hollingsworth}

\affiliation{Materials Physics \& Applications Division: Center for Integrated Nanotechnologies, Los Alamos National Laboratory,
Los Alamos, New Mexico 87545, USA}

\author{Ronen Rapaport}
\affiliation{Racah Institute of Physics, The Hebrew University of Jerusalem, Jerusalem 9190401,Israel}
 \homepage{http://old.phys.huji.ac.il/~ronenr}

\date{\today}
             
\def\thefootnote{$\dagger$}\footnotetext{These authors contributed equally to this work.}\def\thefootnote{\arabic{footnote}}

\begin{abstract}
To harness the potential of radially polarized single photons in applications such as high-dimensional quantum key distribution (HD-QKD) and quantum communication, we demonstrate an on-chip, room-temperature device, which generates highly directional radially polarized photons at very high rates. The photons are emitted from a giant CdSe/CdS colloidal quantum dot (gQD) accurately positioned at the tip of a metal nanocone centered inside a hybrid metal-dielectric bullseye antenna.  We show that due to the large and selective Purcell enhancement specifically for the out-of-plane optical dipole of the gQD, the emitted photons can have a very high degree of radial polarization (>93\%), based on a quantitative metric.
Our study emphasizes the importance of accurate gQD positioning for optimal radial polarization purity through extensive experiments and simulations, which contribute to the fundamental understanding of radial polarization in nanostructured devices and pave the way for implementation of such systems in practical applications using structured quantum light.

\end{abstract}

\maketitle

\section{\label{sec:intro}Introduction}
High-Dimensional Quantum Key Distribution (HD-QKD) extends the principles of photonic-based quantum key distribution beyond traditional binary systems such as standard QKD protocols\cite{hdqkd1,hdqkd2}. A distinct feature of HD-QKD is the use of qudits, which provide additional encoding possibilities, by enabling transmission of more information per quantum state. This expanded information capacity per photon not only facilitates higher key rates but also contributes to increased security by offering resistance to certain eavesdropping attacks. The incorporation of higher-dimensional quantum states thus represents a promising avenue for increasing the rate and security of quantum communication protocols.
The various accessible degrees of freedom of photons, such as spatial\cite{QKDrad3,QKDrad11, QKDrad111}, time-binning\cite{QKD_time}, and vector field polarization\cite{QKDPol} states offer multiple avenues for possible practical realization of HD-QKD.

Radially polarized light is a special case of a family of vector field optical polarization states, where at any point in a plane perpendicular to the optical axis  of the light beam, the electric field vector is pointing in a radial axis emanating from the center of the beam \cite{all_rad_app,rad2}. Radial polarization states offer numerous approaches for HD-QKD protocols, such as multi-array encoding techniques\cite{QKDrad1} and self-healing QKD schemes\cite{QKDrad2}. Notably, photons with radial polarization have been shown as a source for ultrafast encoding on a HD basis of hybrid spatial-polarization modes, requiring only a few simple optical elements that can be modulated very fast with current technologies \cite{hdqkd3}.

Radially polarized photons can in principle be approximately generated from linearly polarized photons using complex phase elements\cite{Qplate,RadBrew,RadG3,RadINter,Radslm,splate2,RadEO, RadFiber, RadFiber2, radMath,waveplate1}. However, such elements introduce unavoidable losses due to residual absorption and scattering, they have to be specially designed to match a given source, and they are bulky and hard to integrate in a compact way, especially with multiple single photon sources (SPS) on a chip, which are considered to be the best sources for QKD systems\cite{sps1,sps2,sps3, sps4}.
Therefore, the development of a compact, on-chip quantum light source emitting high-quality radially polarized photons at ambient conditions\cite{Qinteg, AlexBoaz} can open up opportunities for the implementation of HD-QKD based on vector field polarization\cite{hdqkd4}.

\begin{figure*}
    \includegraphics[width=0.85\textwidth,height=8cm]{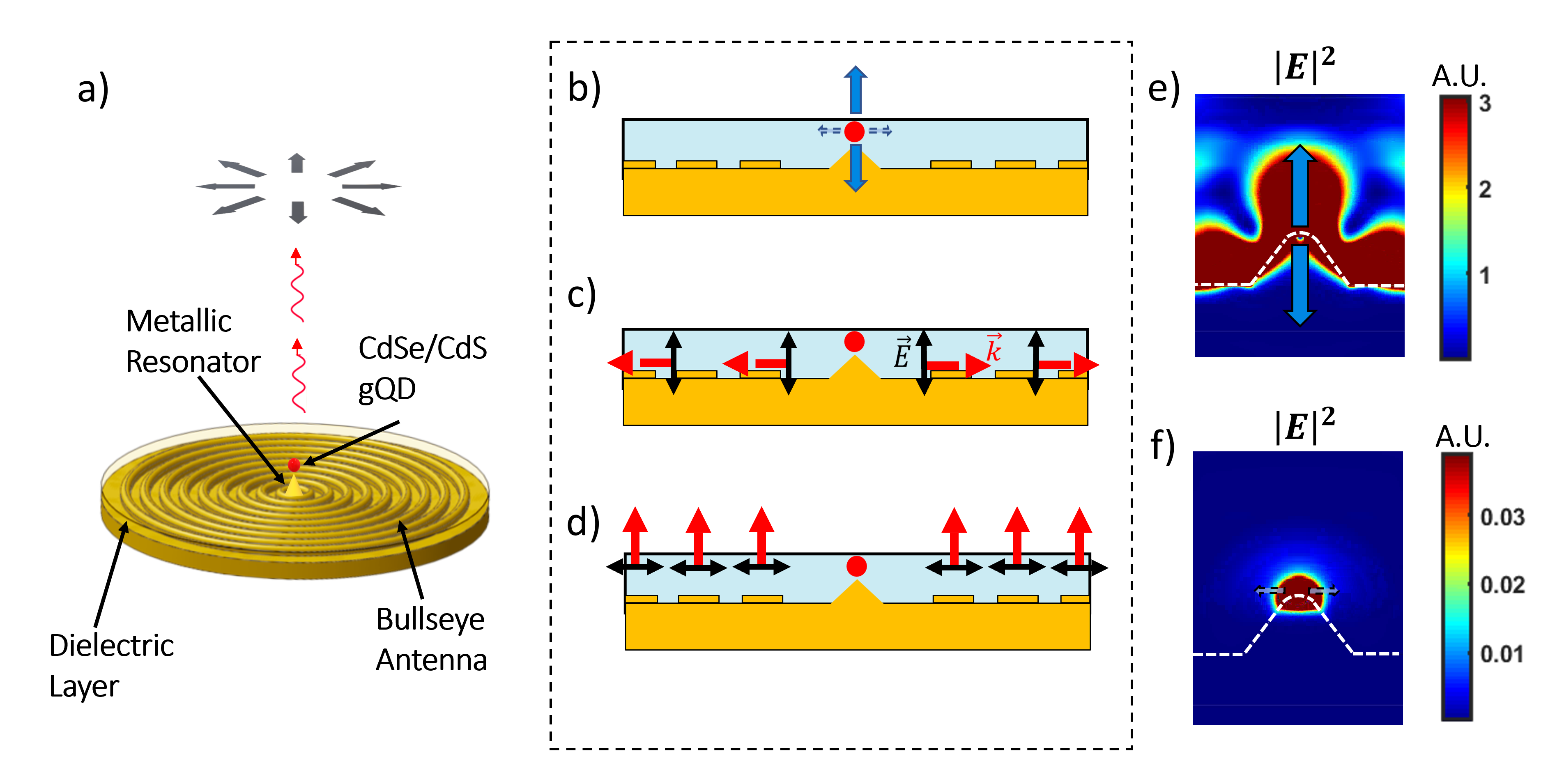}
    \caption{
\textbf{Generation of radially polarized photons - concept.} (a) A schematics of the device. (b)-(d) The concept of the device operation. The nanocone resonantly enhances the out-of-plane dipole emission of the exciton in the gQD (blue) but not the in-plane component (dashed azure). The photon field propagates radially in the dielectric slab waveguide (red arrows in (c)). The circular metallic grating is designed to diffract the radially propagating field by 90 degrees, resulting in a collimated photon field which is radially polarized (d). Near-field FDTD simulations of the out-of-plane dipole radiation (e) and of the in-plane dipole (f), confirming the strong enhancement of the out-of-plane dipole radiation. The colorbars are of intensity counts.}
    \label{fig:fig1}
\end{figure*}

Here we show that an ultrabright, highly collimated on-chip room-temperature SPS, previously demonstrated by our group\cite{AcsHamza,IopHamza,AplHamza}, emits photons with a very high degree of radial polarization. The SPS device is based on a CdSe/CdS gQD\cite{Colloidal_QDS} coupled to a gold nanocone resonator, enclosed within a hybrid metal-dielectric circular Bragg antenna \cite{AcsHamza}. We demonstrate that the emittance of radially polarized photons is achieved through the Purcell enhancement of the out-of-plane dipole, and quantitatively assess the degree of radial polarization (DORP), using a measure based on the axial variation of the horizontal and vertical components of the in-plane polarization. 

In addition, we experimentally establish a direct correlation between the DORP and the precise positioning of the quantum dot at the tip of the nanocone resonator, which is also deduced through Finite Difference Time Domain (FDTD) simulations and is substantiated by comparing results across multiple devices.

\section{Device structure and performance}
Our devices, illustrated in Fig. \ref{fig:fig1}(a) (see SI for details), consist of a CdSe/CdS gQD \cite{Colloidal_QDS} coupled to a metal-dielectric nanoantenna. As illustrated in Fig. \ref{fig:fig1}(b) and discussed in Ref. \cite{AcsHamza}, the surface plasmon resonance of the nanocone at the nanoantenna's center selectively enhances the out-of-plane radiative dipole of the gQD through the Purcell effect\cite{nanocone1, nanocone2, nanocone3, Radant}. The out-of-plane polarized photons, resulting from the Purcell enhanced excitonic recombination of the excited gQD, are emitted preferentially in the radial direction into the dielectric waveguide deposited on top of the gold nanoantenna, as seen in Fig. \ref{fig:fig1}(c). The underlying periodic bullseye structure is designed to serve as a first-order diffraction grating for the wavelength range of the gQD, rotating the propagation direction of the radially propagating emitted photons by 90 degrees, as depicted in Fig. \ref{fig:fig1}(d). This circular grating diffraction is designed such that the outgoing light is highly collimated, for maximal collection efficiency even with low NAs \cite{Livneh2016, AplHamza, AcsHamza}. Importantly, the circular symmetry of the diffraction grating results in redirecting the radially polarized light in the far-field, for an out-of-plane dipole at the center of the antenna.

Near-field FDTD simulations, presented in Fig. \ref{fig:fig1}(e)-(f) (see SI for more details about the simulations), confirm the significant field enhancement of the out-of-plane mode of a radiating dipole, as compared to the in-plane mode. Importantly, the plasmonic resonance can be tuned across optical frequencies by adjusting the nanocone's apex angle to match the gQD's emission properties\cite{IopHamza}.

\begin{figure*}
\includegraphics[width=1.65\columnwidth,height=11.8cm]{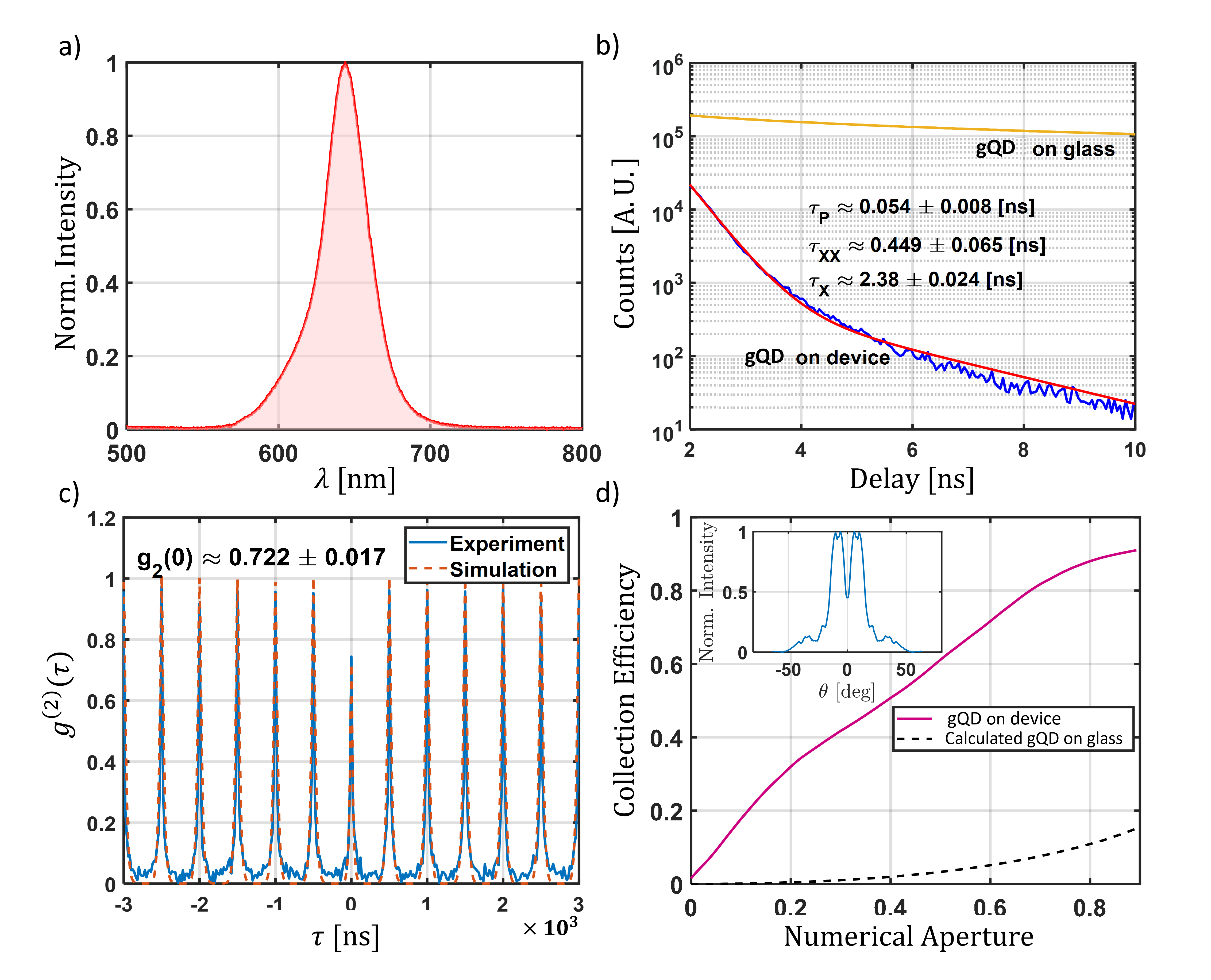}
    \caption{\textbf{Characterization of a typical high-rate, high-directionality photonic device.} (a) Emission spectrum showing the typical CdSe/CdS gQD spectrum at room temperature, with a central wavelength at $\sim 645$ nm\cite{nanocone3}. (b) Lifetime measurements (blue), and a 3-exponent fit for the biexciton, exciton and metal emission\cite{AcsHamza}. The extracted biexciton and exciton lifetimes are $\tau_{XX}=0.449 \pm 0.065$ ns and $\tau_{X}=2.38 \pm 0.024$ ns, respectively; and the plasmon lifetime due to the metal emission is defined as $\tau_{P}$. A lifetime measurement of a single gQD on glass (orange), with extracted biexciton and exciton lifetimes of $\tau_{XX_{REF}}=3.03 \pm 0.31$ ns and $\tau_{X_{REF}}=22.56 \pm 1.20$ ns respectively, is shown as a comparison. The ratio $\tau_{X_{REF}}/\tau_X$ suggests a Purcell enhancement of $P_{F}\sim 10$. (c) Second-order correlation measurements (blue), showing anti-bunching behavior with $g_{2}(0)=0.722\pm 0.017$. The orange dashed line shows a simulated $g_{2}(\tau)$ for a single gQD with both an X and a XX emission (see SI). (d) Photon collection efficiency (CE) from the device (purple) as a function of numerical aperture, compared to a calculated bare gQD on glass (dashed black). An angular profile of the emission is shown in the inset.}
    \label{fig:Fig2}
\end{figure*}

\begin{figure*}
    \includegraphics[width=\linewidth,height=11.5cm]{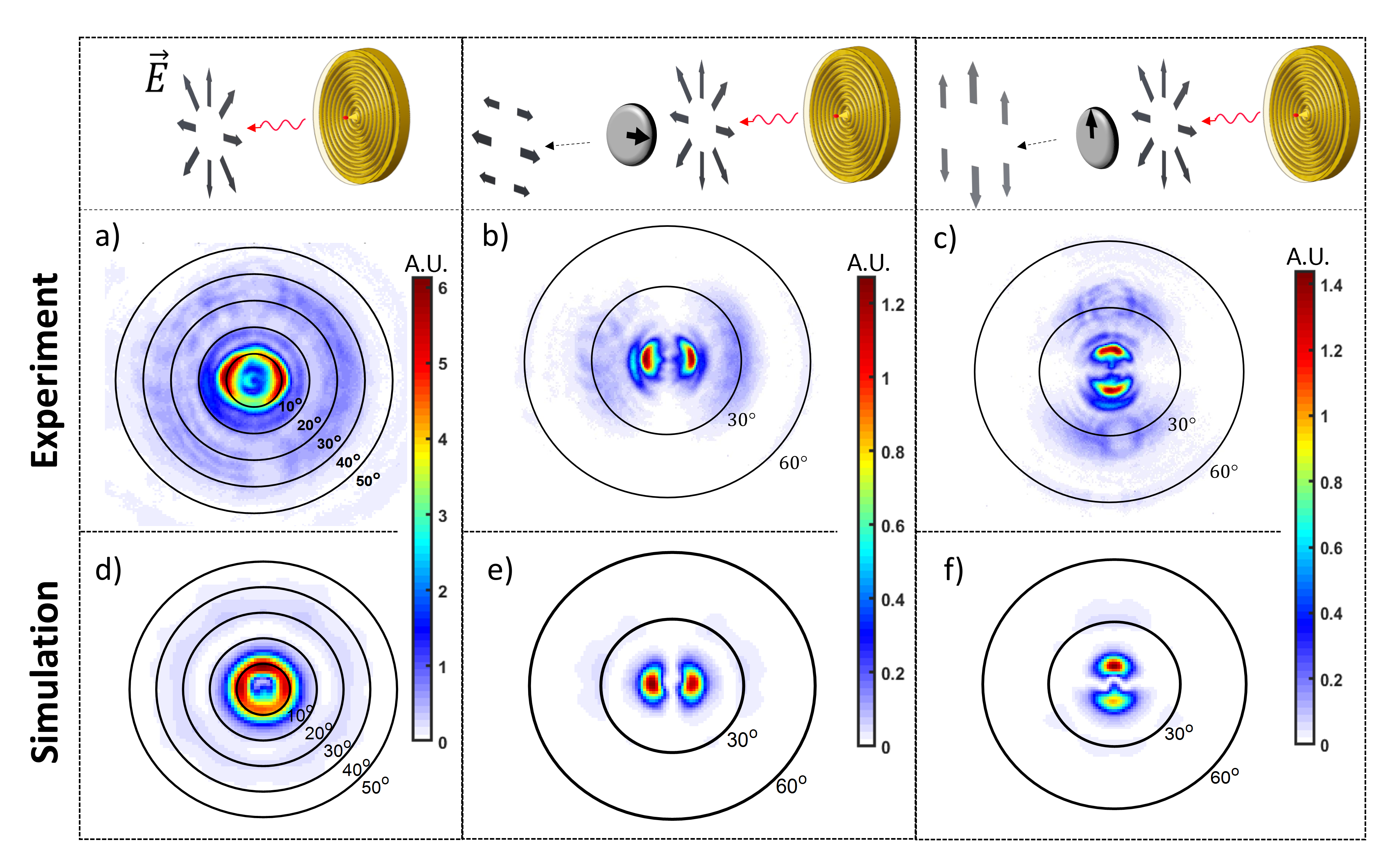}
    \caption{\textbf{Demonstration of radial polarization.} Back focal plane images of the device from Fig. \ref{fig:Fig2} before (a) and after (b),(c) a linear polarizer, compared to corresponding FDTD numerical simulations (d)-(f), displaying the behavior expected from radially polarized light\cite{radMath, Andersen2018}. The polarization orientation is shown in the panel above. The colorbars are of intensity counts.}
    \label{fig:Fig3}
\end{figure*}

Fig. \ref{fig:Fig2}(a)-(b) presents the measured performance of a typical device, including its emission spectrum and the lifetimes of the biexciton and exciton cascades, both significantly shortened compared to a free-standing gQD due to the Purcell factor of the nanocone (see SI for the typical properties of a single gQD). Fig. \ref{fig:Fig2}(c)-(d) shows photon correlations displaying sub-Poissonian light (limited by the residual biexciton emission, see SI), and a high collection efficiency at the first lens due to the excellent collimation effect of the bullseye metal-dielectric antenna. These aspects of the device are similar to the devices previously reported\cite{AcsHamza}.

\begin{figure*}
    \includegraphics[width=0.95\linewidth,height=9.2cm]{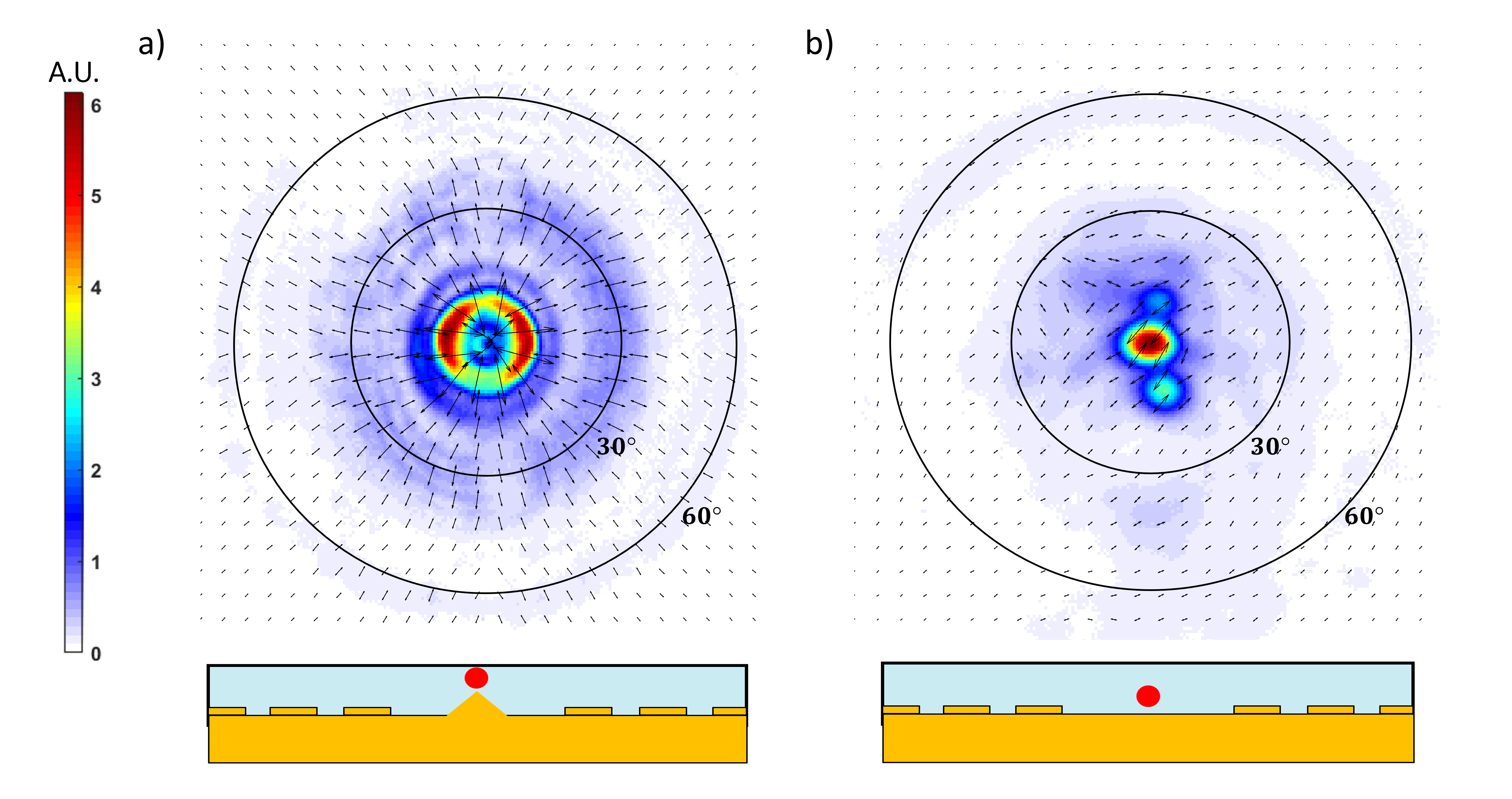}
    \caption{\textbf{Vector field polarization extraction.} The extracted far field electric field vectors of two devices, one with a nanocone (a), showing a clear radial polarization,  and one without the nanocone (b), displaying random polarization. The colorbar is of intensity counts.}
    \label{fig:Fig4}
\end{figure*}

\section{Results and Discussion}

\subsection{Demonstration of Radial Polarization}
Here, we analyze the vectorial polarization properties of the emitted photons. Doing so, we are able to verify that indeed the photons are radially polarized. Specifically, we image the Fourier plane (K-space) of the emission from the device, which gives the far-field angular intensity distribution of the emitted photons. 

Fig. \ref{fig:Fig3}(a) exemplarily shows such a measurement on the device from Fig. \ref{fig:Fig2}. A slowly diverging narrow ring-like angular emission pattern is clearly observed, with a null at its center. This ring-like pattern is strongly indicative of a far-field radial mode of the photons\cite{radMath}.

To verify the radial polarization, we present similar measurements with a linear polarizer (LP) placed after the first lens. As is depicted in the top panel of Fig. \ref{fig:Fig3}, the LP only allows one polarization direction to pass. For a radial field distribution, this should result in two lobes, cut out of the original ring-like pattern, with their connecting axis always along the LP axis, so the lobes should rotate with the LP axis. Images with the LP aligned horizontally and vertically, shown in Fig. \ref{fig:Fig3}(b)-(c), indeed confirm that the rotating lobes are synchronized with the axis of the LP, validating that the original far-field mode has a radial polarization\cite{radMath, Andersen2018}.
We note that Fig. \ref{fig:Fig3}(a)-(c) shows slightly asymmetrical modes, potentially due to emitter placement inaccuracy. This will be discussed later in detail.  

Fig. \ref{fig:Fig3}(d)-(f) presents the far-field FDTD simulations of the device (see SI), with a slight horizontal offset of $\sim$ 5 nm of the emitter from the center of the tip, in order to simulate the slight asymmetry of the experimental results. A very good agreement is seen between experiments and simulations, showing both the radially polarized nature of the emission, as well as the sensitivity to the position of the emitter on the nanocone.

To further ascertain the radial polarization character of the measured device, we deduced the electric field polarization vector at each point of the angular emission image. This was done by analyzing the intensity of each point (each made of a sum over an 8-pixel domain) under 4 different, non-orthogonal directions of the LP axis, as is detailed in the SI. The arrows in Fig. \ref{fig:Fig4}(a) delineate the average orientation (arrow direction) and magnitude (arrow length) of the electric field of the emission, clearly demonstrating radial polarization along the plane. In contrast, Fig. \ref{fig:Fig4}(b) shows a similar analysis on a device consisting of a gQD at the center of a metal-dielectric bullseye antenna but without a nanocone \cite{IopHamza,Livneh2016,AplHamza}. This device shows a highly collimated, Gaussian-like angular emission pattern (unlike the ring emission shown before) with essentially random field polarization. Indeed it was shown before that in such a device  the dominant emission results from the in-plane dipole orientation of the gQD \cite{AplHamza,IopHamza}. Since the in-plane emission has no preferred direction due to the circular symmetry of the antenna, each emitted photon will be randomly polarized.

\begin{figure*}
    \includegraphics[width=1\linewidth,height=15cm]{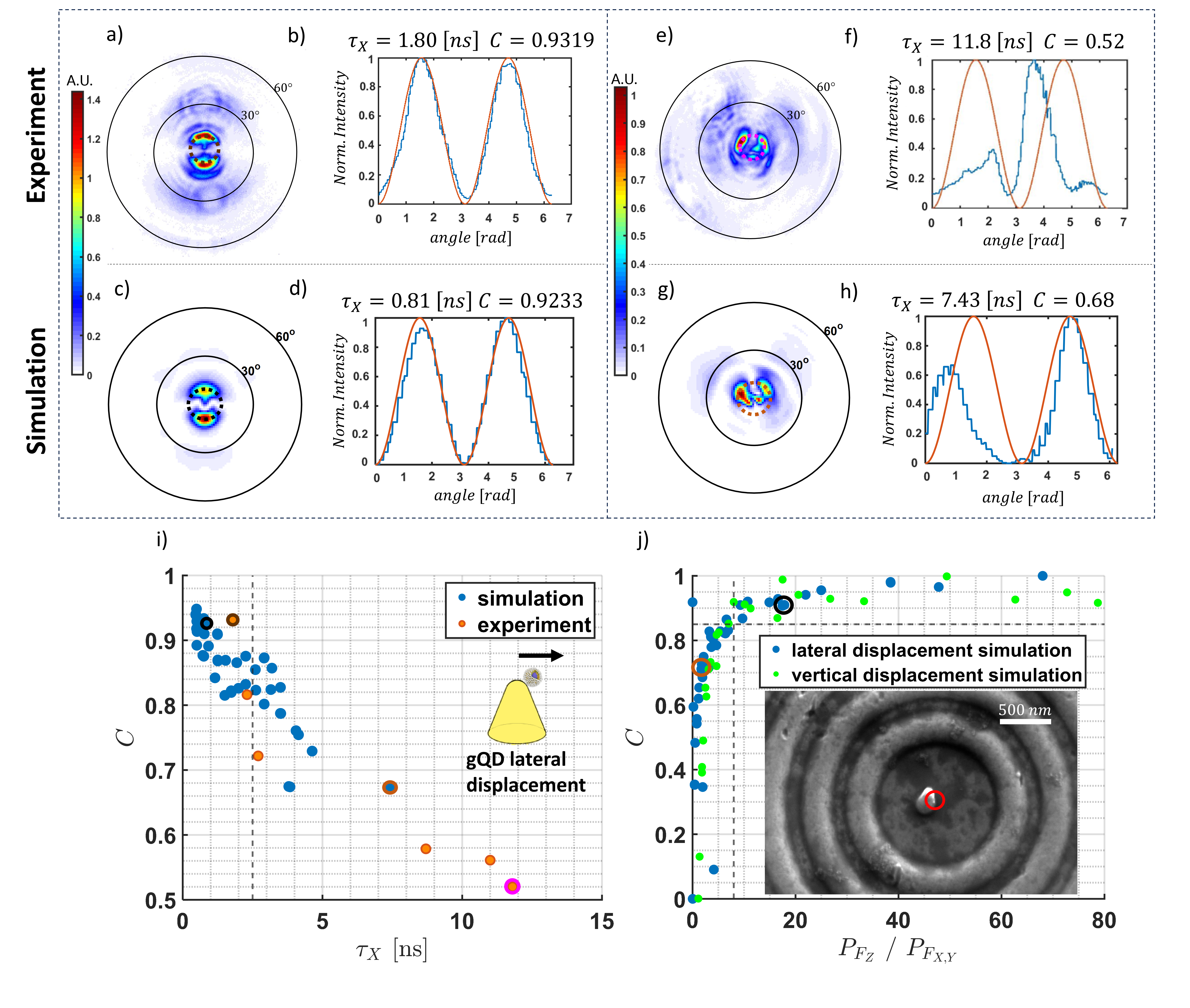}
    \caption{\textbf{Analysis of the degree of radial polarization and its underlying physical mechanism.} (a) Experimental back focal plane image after a horizontal LP of the emission from a device with a short $\tau_{X}=1.8$ ns and a high DORP ($C\simeq0.93$) and (b) the angular intensity profile (blue) fitted to $sin^2(\theta)$ (red). (c),(d) A comparison to a simulated image and intensity profile of a well-positioned emitter, displaced by 5 nm. (e),(f) is similar to (a),(b), but for a device with a long $\tau_{X}=11.8$ ns and a low DORP ($C\simeq0.52$). The simulation in (g),(h) is done with a poorly positioned gQD with a lateral tip displacement of $\sim$40 nm. (i) Correlation between the exciton lifetime $\tau_{X}$ and DORP ($C$). The dashed black line marks a $\sim$ 10 nm tip displacement. Inset is a sketch of the gQD lateral displacement. (j) Plot of simulated Purcell enhancement ratio, $P_{F_Z}/P_{F_{X,Y}}$ and the DORP, for various emitter displacements - vertical (green dots) and lateral (blue dots). The dashed black lines mark the threshold $C$ over which the Purcell factor ratio increases fast. Circle colors match the devices from the top panel. Inset is a SEM image of a device with a displaced gQD (red). The colorbars are of intensity counts.}
    \label{fig:Fig5}
\end{figure*}

\subsection{Degree of Radial Polarization}

After verifying that our devices can indeed emit highly collimated, radially polarized photons at high rates (high Purcell factors), a quantitative way to asses how pure the radial polarization of the photons, and what determines the quality of the radial polarization, is essential. In particular, we found that not all devices perform equally well, both in terms of photon emission rate and the quality of the radial polarization of the emission, and that these two quantities are, in fact, correlated. 

To quantify our discussion we first note that the angular intensity profile of radially polarized light can be expressed as:
\begin{equation}\label{fig:radial_pol}
    I{(r,\theta)} =\left| \begin{pmatrix}
    E_x(r,\theta)\\E_y(r,\theta)
    \end{pmatrix}\right| ^{2}=\left| E(r) \begin{pmatrix}
    cos(\theta)\\sin(\theta)
    \end{pmatrix}\right| ^{2}
\end{equation} 
After passing through a LP having its polarization axis along $\hat{x}$ we have:
\begin{equation}
    \label{eq:radpolLP}
    \begin{aligned}
   I{(r,\theta)} =\left| E_x(r,\theta)\right| ^{2}= 
    \left| E(r) \
    cos(\theta)
    \right| ^{2}
 \end{aligned}
\end{equation}

As can be seen, the normalized intensity at a given radius $r$ from the center is given by, $I_R(\theta)=I_r({\theta})/max(I_r({\theta}))= cos^2(\theta)$, so one can define the DORP for an intensity distribution passing the LP:
\begin{equation}
    \label{eq:C}
    \begin{aligned}
 C &=1 - \sqrt{\frac{4}{3\pi}\int_{\theta}(I_R(\theta) - I_M(\theta))^2 d\theta}
 \end{aligned}
\end{equation}
where $I_M(\theta)=I_m({\theta})/max(I_m({\theta}))$ is the normalized intensity of a general mixed polarization state. For a perfectly radial polarization, $I_M(\theta)=I_R({\theta})$, we have
 $C=1$, while $C<1$ for an imperfect radial polarization. For example, for a random polarization we get $C=0$. 

To quantify $C$ experimentally we define:

\begin{equation}\label{eq:2}\begin{aligned}
C &= 1 - \sqrt{\frac{\Sigma_n(I_R(\theta_n) - I(\theta_n))^2}{\Sigma_n(I_R(\theta_n))^2}}\\
\end{aligned}
\end{equation}
where $n$ denotes an index of a point on the chosen radius and $I(\theta)$ is the normalized measured angular intensity profile of the device.

Fig. \ref{fig:Fig5}(a) shows the back focal plane image of the emission from a device after passing an LP. This device has a considerably shorter exciton lifetime of $\tau_X=1.8$ ns as compared to a free-standing, similar gQD. This short lifetime indicates a large Purcell factor, which for our devices is predominantly for the out-of-plane dipole orientation, as explained above. The normalized angular intensity profile (dashed line) in Fig. \ref{fig:Fig5}(b) is fitted to the theoretical profile of an ideal radial polarization (i.e., $cos^2(\theta)$). As can be seen, a good agreement is found. Using Eq. \ref{eq:2} we indeed find $C=0.9319$, which indicates a very pure radially polarized emission. Fig. \ref{fig:Fig5}(c) shows a simulation with an emitter only slightly displaced horizontally (by $\sim$ 5 nm) from the ideal position above the tip of the nanocone. A similar, slightly shorter lifetime ($\tau_X=0.81$ ns) is found, together with a similar emission profile and similarly high DORP of $C=0.9233$. These results demonstrate the ability of our devices to produce highly collimated, radially polarized single photons, at room temperature. The short, $\sim 1$ns lifetime demonstrates the capability of such devices to emit single photons at rates approaching GHz \cite{AcsHamza}.

In contrast, Fig. \ref{fig:Fig5}(e)-(f) shows the emission profile and angular profile of a device with a much longer lifetime (11.8 ns), which is close to that of the bare gQD. This indicates a poor Purcell factor, likely due to a large displacement of the emitter with respect to the ideal position on the nanocone. A clear deviation from the radial polarization profile is observed.  As expected, the extracted DORP results in a significantly lower value ($C=0.52$). A simulated device with bad emitter positioning (with a displacement of $\sim$ 40 nm) yields a long lifetime of 7.43 ns and a low DORP of only $C=0.68$, as is shown in Fig. \ref{fig:Fig5}(g)-(h).

It thus seems that the DORP and the emission lifetime are closely related. To substantiate this relation, DORP values together with exciton lifetimes, $\tau_X$, where extracted from a series of simulations of displaced emitters, with up to a 40 nm displacement, and are plotted against each other as blue dots in Fig. \ref{fig:Fig5}(i). Experimentally, we extracted the DORP and $\tau_X$ from several devices, which we plot as orange dots on the same figure. A clear correlation is seen for both experimental and simulated data: The shorter $\tau_X$ is, which means the emitter experiences a larger Purcell factor, the larger the DORP. For the shortest $\tau_X$'s, $C\simeq 1$. In contrast, longer $\tau_X$'s, approaching the bare gQD $\tau_X$ (no Purcell radiative enhancement), show very low $C$ values.

The natural explanation for such a distinct relation is through the radiative enhancement of emission via the nanocone-induced Purcell factor. As was discussed above, the nanocone mostly enhances the emission of the out-of-plane dipole. This emission results in radial polarization of the photons in the far field. There is an optimum positioning of the emitter above the tip of the nanocone that yields the largest Purcell factor, and thus the shortest $\tau_X$ and the largest $C$. While the devices are fabricated to meet this geometry, the emitter placement procedure \cite{AcsHamza, nanocone1} allows for small spatial deviations (of a few nanometers) from this optimum position, both laterally and vertically. Such a displacement is seen in the SEM image in the inset of Fig. \ref{fig:Fig5}(j). Such a deviation reduces the Purcell factor of the out-of-plane dipole ($P_{F_{Z}}$), responsible for radial polarization, with respect to that of the in-plane dipole ($P_{F_{X,Y}}$), which is responsible for emission with random polarization. This reduction yields both lifetime increase and $C$ reduction. We note that in contrast, a displacement of the whole nanocone from the center of the cavity only slightly affect the direction of the emission but not its polarization. This was confirmed in our simulations.

To confirm this picture, a comprehensive simulation study explored the influence of emitter displacement on the DORP, considering both vertical and lateral displacements, depicted in Fig. \ref{fig:Fig5}(j). As expected, the larger the ratio $P_{F_{Z}}/P_{F_{X,Y}}$ is, the larger $C$ becomes. Interestingly, $C$ increases fast, and for $P_{F_{Z}}/P_{F_{X,Y}}>10$ the devices have a high DORP with $C>0.85$.
For our devices, this happens at displacements of $\simeq$ 10 nm.

\section{Conclusions}

We demonstrated an on-chip device architecture, based on a gQD coupled to a metal-dielectric nanoantenna, capable of generating single photons with high brightness, high collection efficiency, and with a high degree of a special vectorial polarization state, namely radially polarized photons, at room temperature. Our study revealed, in a quantitative manner, the mechanism behind the purity of the radial polarization,  determined by the positioning of the emitter with respect to its optimal position where the radiative out-of-plane dipole, responsible for the radial polarization, is maximally enhanced.

Overall, these findings validate our integrated compact device concept for an ultrafast and deterministic radial polarization control at the single photon level. Radial polarization has been shown to be a promising source for HD-QKD bases \cite{hdqkd3, hdqkd4}, that can be encoded even at GHz rates using  fast electro-optical modulators \cite{eom}. Our ultrafast room-temperature SPS device \cite{AcsHamza} can thus be a natural source for such demonstrations. This work also opens possibilities for other structured light applications using vector fields polarization with further engineering of the far-field pattern using modified nanocone symmetries.

\begin{acknowledgments}
The gQD synthesis was performed at the Center for Integrated Nanotechnologies (CINT), a Nanoscale Science Research Center and User Facility operated for the U.S. Department of Energy (DOE) Office of Science. The fabrication and characterization of the nano-antenna devices were performed at the Hebrew University center for Nanoscience and Nanotechnology. We thank Saptarshi Ghosh for the SEM images. R.R., Y.B, B.L., H.A., Y.S., and A.N. acknowledge support from the Quantum Communication consortium of the Israeli Innovation Authority.

\end{acknowledgments}

\bibliographystyle{ieeetr}
\bibliography{references}

\end{document}